\documentclass[journal]{IEEEtran}

\usepackage{amsmath,amsfonts,amssymb}
\usepackage{bm}

\usepackage{graphicx}
\usepackage[caption=false,font=footnotesize]{subfig}

\usepackage{cite}

\usepackage{algorithm}
\usepackage{algpseudocode}

\usepackage{tikz}
\usetikzlibrary{
    shapes.geometric,
    arrows.meta,
    positioning
}

\usepackage{enumitem}

\title{Pinching Antenna and Relay Assisted Joint Indoor Outdoor NOMA Communications with SWIPT}
\author{Amit Agarwal, \textit{Member, IEEE,} Susmita Sain,   Keshav Singh, \textit{Senior Member, IEEE},  Sudip Biswas, \textit{Senior Member, IEEE}
\thanks{Amit Agarwal is an Assistant Professor at ECE department, VNIT Nagpur, India. Keshav Singh is associated with National Sun Yat-sen University,Taiwan.  Susmita and Sudip is asscociate with IIIT Guwahai, India.  Email: amit.nitk07@gmail.com, susmita.sain23b@iiitg.ac.in, keshav.singh@mail.nsysu.edu.tw, sudip.biswas@iiitg.ac.in.}}

\begin{document}
\maketitle
\begin{abstract}
This paper presents a unified downlink architecture for serving indoor and outdoor users from a shared multi-antenna rooftop base station (BS). To overcome severe building-penetration loss, the BS transfers information and radio-frequency (RF) energy to an indoor access point (AP) using simultaneous wireless information and power transfer (SWIPT). The AP harvests this energy and performs amplify-and-forward (AF) relaying, enabling self-sustained indoor communication. A movable waveguide-mounted pinching antenna (PA) further improves indoor coverage by adapting its position to the propagation environment. For outdoor transmission, zero-forcing (ZF) beamforming suppresses inter-group interference, while power-domain non-orthogonal multiple access (NOMA) supports multiple users within each group. We maximize the outdoor weighted sum rate (WSR) while satisfying stringent indoor and outdoor quality-of-service (QoS) requirements. The resulting nonconvex problem jointly optimizes the BS power fraction, SWIPT power-splitting ratio, PA position, and NOMA power allocation. To obtain an efficient solution, we develop a two-stage method based on alternating optimization (AO) and sequential quadratic programming (SQP). Numerical results show that the proposed architecture improves outdoor spectral efficiency while maintaining reliable, energy-sustainable indoor coverage. Increasing the BS array size further raises the indoor QoS-satisfaction probability toward unity, even under demanding rate targets.

\end{abstract}

\section{Introduction}
Sixth-generation (6G) wireless networks aim to provide highly spectral- and energy-efficient communication for both indoor and outdoor users. The use of higher-frequency bands, such as millimeter-wave (mmWave) and terahertz (THz), however, makes wireless links more sensitive to blockage and path loss. In addition, indoor users experience severe penetration loss through building walls, making direct transmission from an outdoor base station (BS) less reliable \cite{tvt1}. Relaying provides an effective solution to this problem. Among different relaying strategies, amplify-and-forward (AF) relaying is attractive due to its low processing complexity and latency \cite{a2,a4}. However, conventional relays require an external power supply, which increases deployment complexity. Simultaneous wireless information and power transfer (SWIPT) addresses this limitation by enabling the relay to harvest energy from the received RF signal while processing information \cite{s1}, thereby enabling self-powered AF relaying.

Pinching antennas mounted on dielectric waveguides provide another degree of freedom for improving indoor communication. By flexibly adjusting the radiation position according to user locations, the resulting propagation distance and channel gain can be optimized \cite{c1,c2}. For outdoor users, power-domain non-orthogonal multiple access (NOMA) enables multiple users to share the same time--frequency resource through power-domain multiplexing and successive interference cancellation (SIC) \cite{noma1,noma2}. However, when indoor and outdoor users are simultaneously served by the same BS, interference between the two transmissions can degrade system performance. Zero-forcing (ZF) beamforming can mitigate this interference by spatially separating the indoor and outdoor transmissions.

Although SWIPT relaying, pinching antennas, and NOMA have been extensively studied individually, their joint integration introduces significant optimization challenges. The BS power allocation factor affects both the harvested relay energy and the power available to outdoor NOMA users. Meanwhile, the SWIPT power-splitting ratio and pinching-antenna position determine the indoor achievable rate, while the outdoor NOMA power coefficients are coupled through SIC interference and QoS constraints. Therefore, the resulting joint resource-allocation problem is inherently non-convex.

To address these challenges, this paper proposes a joint indoor--outdoor communication framework integrating a SWIPT-powered AF relay, pinching antenna, ZF beamforming, and outdoor NOMA. The proposed approach first optimizes the indoor operating point, including the BS power allocation, SWIPT power-splitting ratio, and pinching-antenna position, using alternating optimization (AO). The remaining BS power is then allocated among outdoor NOMA users using sequential quadratic programming (SQP) subject to individual QoS requirements. This decomposition enables efficient coordination between indoor coverage and outdoor spectral efficiency.

The main contributions of this work are summarized as follows:
\begin{itemize}

    \item We introduce a unified indoor-outdoor transmission architecture that synergistically integrates a SWIPT-enabled amplify-and-forward (AF) relay, a movable waveguide-mounted PA, ZF beamforming, and power-domain NOMA. This architecture enables energy-sustainable indoor connectivity under severe building-penetration loss while spatially isolating the indoor and outdoor transmission streams.

    \item We formulate a joint resource-allocation problem that maximizes the outdoor weighted sum rate (WSR) subject to stringent quality-of-service (QoS) requirements for all users. The formulation captures the intricate coupling among the BS indoor-power fraction, the SWIPT power-splitting ratio, the PA position, and the outdoor NOMA power-allocation coefficients.

    \item We develop a two-stage optimization framework tailored to the structure of the resulting nonconvex problem. In the first stage, an alternating optimization (AO) procedure jointly determines the indoor-power fraction, power-splitting ratio, and PA position, thereby establishing an energy-feasible operating point that guarantees the indoor QoS requirements. Numerical evaluations confirm the stable convergence of the proposed procedure.

    \item Using the optimized indoor operating point, we transform the remaining design into a structured outdoor NOMA resource-allocation problem and solve it efficiently through SQP. This stage dynamically distributes the residual BS transmit power according to the users' channel conditions and priority weights, enabling QoS-aware and priority-sensitive multiplexing.

    \item Extensive numerical results reveal the performance gains and operating tradeoffs of the proposed architecture. The outdoor WSR reaches $3.1953$~bps/Hz at a transmit signal-to-noise ratio (SNR) of $110$~dB. Increasing the BS antenna array from $M=8$ to $M=16$ raises the indoor QoS-satisfaction probability to $100\%$ for target rates of up to $1.0$~bps/Hz. The results further demonstrate that equal user weighting is generally suboptimal, highlighting the importance of priority-aware NOMA power allocation for maximizing the outdoor WSR.

\end{itemize}

The remainder of this paper is organized as follows. Section II presents the system model and the considered indoor--outdoor transmission architecture. Section III formulates the joint resource-allocation problem and presents the proposed optimization framework. Section IV provides numerical results and discusses the impact of the system parameters on indoor QoS and outdoor NOMA performance. Finally, Section V concludes the paper.
\begin{figure}[!t]
    \centering
    \includegraphics[width=8cm]{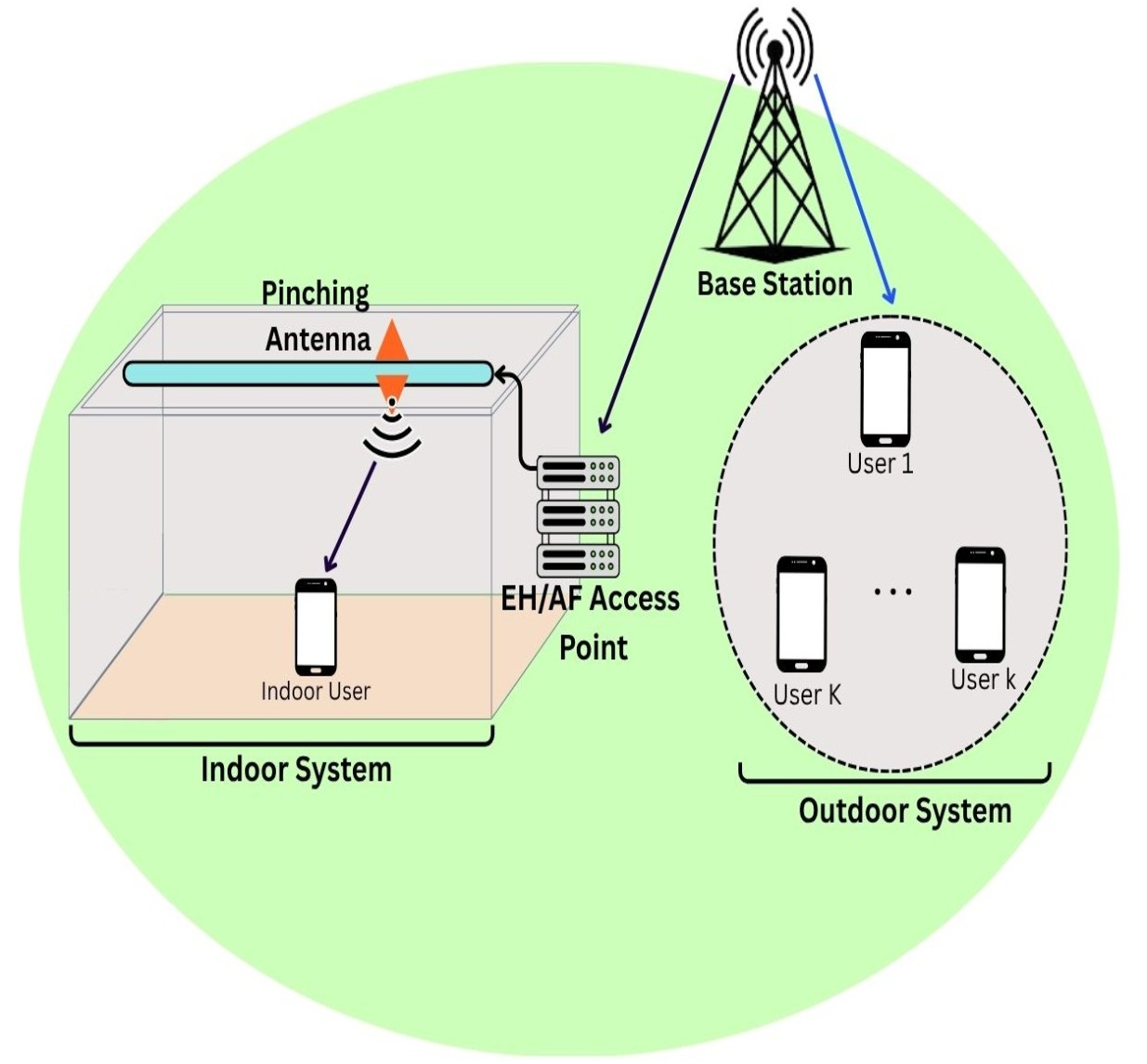}
    \caption{Proposed downlink integrated indoor--outdoor communication system architecture featuring ZF beamforming at the rooftop base station, a self-powered SWIPT-enabled AF relay with a waveguide-mounted pinching antenna. AP is the feed point at $[0,0,3]$ m.}
    \label{fig:SystemModel}
\end{figure}

\section{System Model}
As shown in Fig.~\ref{fig:SystemModel}, we consider a downlink integrated indoor--outdoor communication system. A rooftop base station (BS) with $M$ antennas simultaneously serves single indoor and multiple outdoor users via hybrid space division multiple access (SDMA) and NOMA using total power $P$. Specifically, the BS employs ZF beamforming to separate the indoor and outdoor links in the spatial domain. A fraction $\rho P$ ($\rho\in[0,1]$) is allocated to the indoor link, and $(1-\rho)P$ to the outdoor link. The indoor stream is routed through a self-powered, SWIPT-enabled amplify-and-forward (AF) access point (AP) relay that uses harvested energy to forward signals without decoding. To enhance the data rate, the AP utilizes a pinching antenna (PA) movable along a dielectric waveguide to optimize its radiation point. Outdoor users are served directly by the BS using power-domain NOMA with successive interference cancellation (SIC).

The transmitted signal is expressed as
\begin{equation}
\mathbf{x} = \sqrt{\rho P}\mathbf{w}_{\mathrm{in}} s_{\mathrm{in}} + \sqrt{(1-\rho)P}\mathbf{w}_{\mathrm{out}} s_{\mathrm{out}},
\label{eq:tx}
\end{equation}
where $\mathbf{w}_{\mathrm{in}}, \mathbf{w}_{\mathrm{out}}\in\mathbb{C}^{M\times1}$ are the indoor and outdoor beamforming vectors, respectively. The normalized NOMA symbols satisfy $\mathbb{E}\{|s_{\mathrm{in}}|^2\} = \mathbb{E}\{|s_{\mathrm{out}}|^2\} = 1$. The outdoor symbol is $s_{\mathrm{out}} = \sum_{k=1}^{K_{\mathrm{out}}} \sqrt{b_k}\,x_k$, where $b_k$ denotes the NOMA power allocation coefficients satisfying $\sum_{k=1}^{K_{\mathrm{out}}}b_k=1$ and $b_1\ge b_2\ge\cdots\ge b_{K_{\mathrm{out}}}$. 

\subsection{Channel Model}
The channels are assumed to be constant within each transmission block. 

\subsubsection{BS-to-Relay Channel}
The wireless channel link between the rooftop BS and the indoor AP relay is denoted by the vector $\mathbf{h}_{BR}\in\mathbb{C}^{M\times1}$. This link is mathematically modeled as
\begin{equation}
\mathbf{h}_{BR} = \sqrt{\beta_{BR}} \mathbf{g}_{BR},
\label{eq:h_BR}
\end{equation}
where $\beta_{BR}$ represents the large-scale path-loss attenuation coefficient and $\mathbf{g}_{BR}$ captures the small-scale fading characteristics. The small-scale fading vector follows a complex circularly symmetric Gaussian distribution given by
$\mathbf{g}_{BR} \sim \mathcal{CN}(\mathbf{0}, \mathbf{I}_M),$
where $\mathbf{I}_M$ is the $M \times M$ identity matrix. The large-scale path loss factor is explicitly formulated as
$\beta_{BR} = C_0 d_{BR}^{-\alpha_{BR}},
$
where $C_0$ denotes the reference path loss measured at a distance of one meter, we assume $C_0=1$, $d_{BR}$ represents the physical line-of-sight propagation distance separating the base station from the relay access point, and $\alpha_{BR}$ represents the corresponding large-scale path-loss exponent.

\subsubsection{Relay-to-Indoor User Pinching Channel}
The system serves a single indoor user located at a fixed coordinate vector $\mathbf{u}_{\mathrm{IN}} = [x_{\mathrm{IN}}, y_{\mathrm{IN}}, 0]^T$. The AP acts as a corner-fed feed point ($q=0$) for a horizontal dielectric waveguide at height $h$. The channel $h_{RI}(q)$ between the movable pinching antenna (PA) at position $q$ and the user is modeled as
\begin{equation}
h_{RI}(q) = \frac{\sqrt{\eta}}{d_{\mathrm{IN}}(q)} \exp(-\alpha q) \exp\!\left(-j\frac{2\pi}{\lambda}d_{\mathrm{IN}}(q)\right),
\label{eq:h_RI}
\end{equation}
where $\lambda$ is the wavelength, $\eta=(\lambda/4\pi)^2$ is the free-space propagation constant, and $\alpha$ is the waveguide attenuation coefficient. The term $\exp(-\alpha q)$ scales the internal signal loss from the corner-fed feed point to the radiation point $q$. The physical propagation distance is given by $d_{\mathrm{IN}}(q) = \sqrt{(x_{\mathrm{IN}}-q)^2 + y_{\mathrm{IN}}^2 + h^2}$.

\subsubsection{Outdoor BS-to-User Channel}
The BS directly serves $K$ outdoor users. For user $k$ at distance $d_k$ and angle of departure (AoD) $\theta_k$, the large-scale path loss is $\beta_k = (\frac{\lambda}{4\pi d_k})^2$. The outdoor channel $\mathbf{h}_{ k}$ follows a Rician fading model:
\begin{equation}
\mathbf{h}_{k} = \sqrt{\beta_k} \left( \sqrt{\frac{K_{\mathrm{R}}}{K_{\mathrm{R}}+1}} \mathbf{a}(\theta_k) + \sqrt{\frac{1}{K_{\mathrm{R}}+1}} \mathbf{g}_k \right),
\label{eq:rician_channel}
\end{equation}
where $K_{\mathrm{R}}$ is the Rician factor, $\mathbf{a}(\theta_k)$ is the array steering vector, and $\mathbf{g}_k$ is the non-line-of-sight Rayleigh component.

\section{Problem Formulation and Proposed Solution}
To eliminate mutual interference, the BS employs ZF beamforming satisfying $\mathbf{H}_{\mathrm{out}}^{H}\mathbf{w}_{\mathrm{in}}=\mathbf{0}$ and $\mathbf{h}_{BR}^{H}\mathbf{w}_{\mathrm{out}}=0$, where $\mathbf{H}_{\mathrm{out}}$ is the aggregate outdoor channel matrix. Thus, the received signal at the AP is isolated from outdoor interference and is expressed as
\begin{equation}
y_R = \sqrt{\rho P} \mathbf{h}_{BR}^{H} \mathbf{w}_{\mathrm{in}} s_{\mathrm{in}} + n_R,
\label{eq:relayrx}
\end{equation}
where $n_R\sim\mathcal{CN}(0,\sigma_R^2)$ is the AWGN at the relay.

\begin{algorithm}[!t]
\footnotesize
\caption{Joint Indoor-Outdoor Resource Allocation}
\label{alg:joint}
\begin{algorithmic}[1] 
\raggedright
\State \textbf{Input:} $P, \eta, \alpha, \lambda, \sigma^2, \sigma_R^2, \zeta, R^{\min}_{\rm in},  \gamma_{\rm th}, K_{\rm out}=K, M$.
\State Compute projection matrix $\mathbf{V}=\text{null}(\mathbf{H}_{\rm out})$.
\State Set indoor beamformer $\mathbf{w}_{\rm in} = \frac{\mathbf{V}\mathbf{V}^{H}\mathbf{h}_{BR}}{\|\mathbf{V}\mathbf{V}^{H}\mathbf{h}_{BR}\|}$ and $G_{BR}=|\mathbf{h}_{BR}^{H}\mathbf{w}_{\rm in}|^2$.
\State Initialize $\rho, \rho_R, q$ and set $\rho_{\rm old} = \rho$.
\For{$i=1$ to $I_{\max}$}
    \State \textit{\% Step 1: Minimum BS Power Fraction}
    \For{each feasible $\rho$ in ascending order}
        \If{$R_{\rm in}(\rho, \rho_R, q) \geq R_{\rm in}^{\min}$}{ $\rho^\star \leftarrow \rho$; \textbf{break}} \EndIf
    \EndFor
    \State \textit{\% Step 2: Update PS Ratio}
    \For{each feasible $\rho_R$}
        \If{$R_{\rm in}(\rho^\star, \rho_R, q) \geq R_{\rm in}^{\min}$}{ $\rho_R^\star \leftarrow \rho_R$; \textbf{break}} \EndIf
    \EndFor
    \State \textit{\% Step 3: Optimize PA Position}
    \State Set $R_{\max}=0$.
    \For{each feasible position $q$}
        \If{$R_{\rm in}(\rho^\star, \rho_R^\star, q) \geq R_{\rm in}^{\min}$ and $R_{\rm in}>R_{\max}$}
            \State $R_{\max}\leftarrow R_{\rm in}$, $q^\star\leftarrow q$
        \EndIf
    \EndFor
    \State \textbf{if} $|\rho^\star - \rho_{\rm old}| < 10^{-4}$ \textbf{then break} \textit{\% Convergence Check}
    \State Update reference variable: $\rho_{\rm old} \leftarrow \rho^\star$
\EndFor
\State Compute $R_{\rm in}=\log_2(1+\gamma_{\rm in})$.
\If{$R_{\rm in} \ge R^{\min}_{\rm in}$}
    \State Set $\mathbf{W}_{\rm out} = \frac{\mathbf{V}_{\rm out}(:,1)}{\|\mathbf{V}_{\rm out}(:,1)\|}$ where $\mathbf{V}_{\rm out} = \text{null}(\mathbf{h}_{BR}^{H})$.
    \State Calculate $G_k = |\mathbf{H}_{\rm out}(k,:)\mathbf{W}_{\rm out}|^2$ for all $k \in \{1,\dots, K_{\rm out}\}$.
    \State Sort $\{G_k\}$ in ascending order to establish NOMA SIC chain.
    \State Set remaining outdoor power: $P_{\rm out}=(1-\rho^\star)P$.
    \State Solve $\mathcal{P}(1.2)$ via SQP to obtain NOMA vector $\mathbf{b}^\star$.
\Else
    \State \textbf{Stop:} Target indoor QoS demands cannot be sustained.
\EndIf
\State \textbf{Output:} Optimal $\{\rho^\star, \rho_R^\star, q^\star, \mathbf{b}^\star\}$, $R_{\rm in}$, and vector $\mathbf{R}_{\rm out}$.
\end{algorithmic}
\end{algorithm}

\subsection{SWIPT-Based Energy Harvesting at the Relay}

The indoor AP is equipped with a power-splitting (PS) SWIPT receiver. Upon receiving the RF signal from the BS, a fraction $\rho_R$ of the received power is allocated for information processing, while the remaining fraction $(1-\rho_R)$ is utilized for energy harvesting, where $0\leq\rho_R\leq1$ denotes the power-splitting ratio. Accordingly, the harvested power at the relay is given by
$P_R
=
\xi(1-\rho_R)\rho P
\left|
\mathbf{h}_{BR}^{H}\mathbf{w}_{\rm in}
\right|^2,
$
where $\xi\in(0,1]$ denotes the RF-to-DC energy conversion efficiency. Assuming that all the harvested energy is immediately used for relay transmission, the available relay transmit power is
\begin{equation}
P_R
=
\xi (1-\rho_R)\rho P
\left|
\mathbf{h}_{BR}^{H}\mathbf{w}_{\rm in}
\right|^2.
\label{eq:PR}
\end{equation}
\subsection{AF Relay and Data Rates}
The self-powered relay adopts the AF protocol to retransmit the indoor signal without decoding. To satisfy the transmit power constraint $\mathbb{E}\{|x_R|^2\}\leq P_R$, the relay amplification coefficient $\beta_R$ is obtained as
\begin{equation}
\beta_R = \sqrt{ \frac{P_R}{\rho_R \rho P \left|\mathbf{h}_{BR}^{H}\mathbf{w}_{\rm in}\right|^2 +\sigma_R^2} }.
\label{eq:beta}
\end{equation}
By incorporating the amplified relay transmission into the indoor receiver model, the received SINR at the single indoor user is expressed as
\begin{equation}
\gamma_{\rm in} = \frac{{\rho_R}\rho P \beta_R^2 |h_{RI}(q)|^2 |\mathbf{h}_{BR}^{H}\mathbf{w}_{\rm in}|^2}{\beta_R^2 |h_{RI}(q)|^2 \sigma_R^2 + \sigma^2}.
\label{eq:SINRindoor}
\end{equation}
Accordingly, the corresponding achievable data rate of the indoor user is given by
\begin{equation}
R_{\rm in} = \log_2(1+\gamma_{\rm in}) \quad \text{[bps/Hz]}.
\label{eq:Rin}
\end{equation}
\subsection{Outdoor Data Rate}
For the outdoor subsystem, the base station directly serves $K$ users through power-domain NOMA and SIC decoding based on channel quality ordering. The received SINR for outdoor user $k$ is formulated as
\begin{equation}
\gamma_{{\rm out},k} = \frac{(1-\rho) P b_k |\mathbf{h}_{k}^{H} \mathbf{w}_{\rm out}|^2}{(1-\rho) P |\mathbf{h}_{k}^{H} \mathbf{w}_{\rm out}|^2 \sum_{m=k+1}^{K_{}}b_m +\sigma_k^2},
\label{eq:SINRout}
\end{equation}
where $\sigma_k^2$ denotes the local noise variance. Finally, the individual achievable data rate for outdoor user $k$ is given by
\begin{equation}
R_{{\rm out},k} = \log_2(1+\gamma_{{\rm out},k}) \quad \text{[bps/Hz]}.
\label{eq:Rout}
\end{equation}


\subsection{Problem Formulation}
To maximize the weighted sum rate (WSR) of outdoor users while ensuring QoS, the joint optimization over the indoor power fraction $\rho$, power-splitting ratio $\rho_R$, antenna position $q$, and NOMA vector $\mathbf{b}=[b_1,\ldots,b_{K_{\rm out}}]^T$ is formulated as
\begin{equation}
\label{eq:P1}
\begin{aligned}
\mathcal{P}(1):\ \max_{\rho,\rho_R,q,\mathbf{b}}\ &\sum_{k=1}^{K_{\rm out}} w_k R_{{\rm out},k} \\
\text{s.t.}\ \quad &\text{C}_1: R_{\rm in}(\rho,\rho_R,q)\geq R_{\rm in}^{\min}, \\
&\text{C}_2: R_{{\rm out},k}(\rho,\mathbf{b})\geq R_{{\rm out},k}^{\min}, \ \forall k, \\
&\text{C}_3: \textstyle\sum_{k=1}^{K_{\rm out}}b_k=1, \\
&\text{C}_4: 0\leq b_k \leq1,\ \forall k, \\
&\text{C}_5: 0\leq\rho\leq1,\quad 0\leq\rho_R\leq1, \\
&\text{C}_6: q_{\min}\leq q\leq q_{\max}.
\end{aligned}
\end{equation}

In problem $\mathcal{P}(1)$, the constraints represent the quality-of-service (QoS) and physical resource limitations of the joint framework. Specifically, $\text{C}_1$ and $\text{C}_2$ establish the minimum data rate requirements ($R_{\rm in}^{\min}$ and $R_{{\rm out},k}^{\min}$) needed to guarantee the QoS of the indoor user and each individual outdoor NOMA user, respectively. Constraint $\text{C}_3$ ensures that the total transmission power allocated to the outdoor subsystem is fully distributed among the group users, while $\text{C}_4$ guarantees that each individual NOMA fractional coefficient remains within a valid normalized range. Similarly, $\text{C}_5$ restricts both the base station indoor power allocation fraction $\rho$ and the relay power-splitting ratio $\rho_R$ to the physically feasible bounds of zero and one. Finally, $\text{C}_6$ bounds the pinching-antenna position $q$ to the actual physical length dimensions of the deployed dielectric waveguide ($q_{\min}$ to $q_{\max}$).

Due to coupled variables in the indoor rate and outdoor NOMA terms, non-convex problem $\mathcal{P}(1)$ is split into two successive subproblems. First, we minimize the indoor power fraction to maximize the outdoor power $P_{\rm out}^\star=(1-\rho^\star)P$:
\begin{equation}
\label{eq:P1_1}
\begin{aligned}
\mathcal{P}(1.1):\ \min_{\rho,\rho_R,q}\ &\rho \\
\text{s.t.}\ \quad &\text{C}_1, \text{C}_5, \text{C}_6.
\end{aligned}
\end{equation}
We solve $\mathcal{P}(1.1)$ using an alternating optimization (AO) procedure that sequentially updates $\rho$, $\rho_R$, and $q$ until convergence, yielding $(\rho^\star, \rho_R^\star, q^\star)$.

Second, using the fixed $\rho^\star$, the outdoor NOMA power allocation subproblem is expressed as
\begin{equation}
\label{eq:P1_2}
\begin{aligned}
\mathcal{P}(1.2):\ \max_{\mathbf{b}}\ &\sum_{k=1}^{K_{\rm out}} w_k R_{{\rm out},k}(\rho^\star,\mathbf{b}) \\
\text{s.t.}\ \quad &\text{C}_2, \text{C}_3, \text{C}_4.
\end{aligned}
\end{equation}
Problem $\mathcal{P}(1.2)$ is solved via sequential quadratic programming (SQP), yielding the final solution vector $\{\rho^\star,\rho_R^\star,q^\star,\mathbf{b}^\star\}$ summarized in Algorithm 1.

\subsection{Convergence of $\mathcal{P}(1.1)$}
We assume that the noise at each user is independent with variance $\sigma^2$. The transmit SNR is defined as
$
\frac{P}{\sigma^2}.
$ For a given channel realization with $M=8,16$, $P/\sigma^2=110$ dB,  $K=4$@ [10,12,15,18] m and angle uniformly distributed  $[-\pi,\pi]$, $\mathbf{w}=[0.1,0.4,0.3,0.2]$, user location $\mathbf{u}=[2,5]$, and $R_{\rm in}^{\min}=R_{{\rm out},k}^{\min}={0.5,1}$ bps/Hz, the  algorithm converges to optimum $\rho$ values. Specifically, this  yields an indoor SNR $\rho^* P/\sigma^2$ of $\approx -25$ dB for $M=16$  satisfying the $0.5$  QoS requirement while securing an outdoor NOMA WSR of $1.6281$ bps/Hz.  The corresponding $\rho_R^*=.003,q^\star=1.69$. Due to waveguide attenuation $\alpha$, the  antenna position $q^\star$ offsets from the user's coordinate $x=2$. All other parameters are taken from Table~\ref{tab:simulation_parameters}.

\begin{figure}
    \centering
    \includegraphics[width=8.5cm]{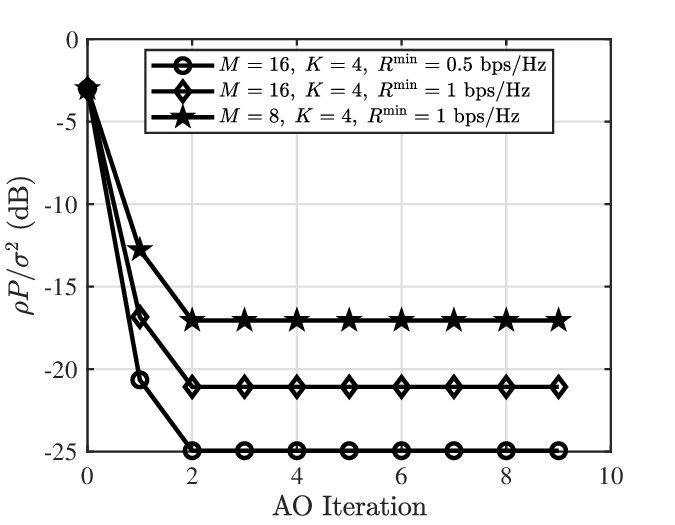}
    \caption{convergence of $\mathcal{P}(1.1)$ with $R_{\rm in}^{\min}=
    R_{{\rm out},k}^{\min}=R^{\min}$.}
    \label{fig:conv}
\end{figure}

\begin{table}[b]
\centering
\caption{Simulation Parameters}
\label{tab:simulation_parameters}
\footnotesize
\begin{tabular}{l c | l c}
\hline
\textbf{Parameter} & \textbf{Value} & \textbf{Parameter} & \textbf{Value} \\ \hline
$K$ & $2, 4$ & $M$ & $8, 16$ \\
$N_{\mathrm{real}}$ & $5000$ & $\mathbf{u}_{\mathrm{IN}}$ & $[2,5,0]^T$\,m \\
$P/\sigma^2$ & $110$\,dB & $\xi$ & $0.9$ \\
$f_c$ & $28$\,GHz & $\alpha$ & $0.009$ \\
$h$ & $3$\,m & $R_{\rm in}^{\min}$ & $0.5$\,bps/Hz \\
$\mathbf{R}_{\rm out,k}^{\min}$ & $0.5$\,bps/Hz & $N_{\mathrm{iter}}$ (AO) & $30$ \\
$q_{\min}, q_{\max}$ & $0, 10$\,m & $\alpha_{BR}$ & $2.2$ \\
$d_{BR}$ & $10$\,m & $K_R$ & $5$\,dB \\ \hline
\end{tabular}
\end{table}

\begin{figure}[!t]
    \centering
    \includegraphics[width=8.5cm]{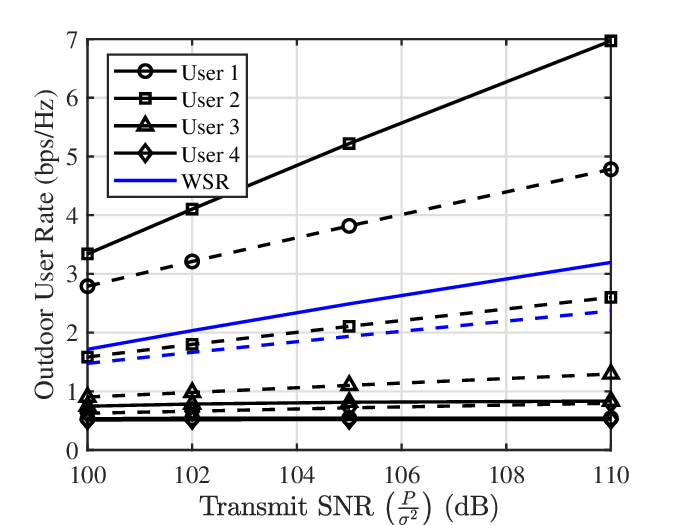}
    \caption{Individual user data rates and outdoor NOMA weighted sum rate versus the transmit SNR up to $110$~dB, $M=16$ for $K_{\rm out}=4$ users. Solid lines correspond to unequal weights, dashed lines correspond to equal weights of $0.25$ each, and the targets are fixed at $R_{\rm in}^{\min}=R_{{\rm out},k}^{\min}=0.5$~bps/Hz.}
    \label{fig:Pt_rate}
\end{figure}

\section{Simulation Results}
We evaluate the performance achieved by the proposed framework. Unless otherwise specified, all system parameters are taken from Table~\ref{tab:simulation_parameters}. All curves represent ergodic performance, with the results averaged over $N_{\rm real}$ independent channel realizations.

Fig.~\ref{fig:Pt_rate} illustrates the outdoor-user rates versus the transmit SNR, where the minimum rate threshold is set to $0.5$~bps/Hz. The four outdoor users, $U_1$ to $U_4$, are located at distances of $10$, $12$, $15$, and $18$ meters, respectively, with angular positions generated uniformly over $[-\pi,\pi]$. Under the unequal weighting vector $\boldsymbol{w}=[0.1, 0.4, 0.3, 0.2]$, User~2 is prioritized, leading its rate to scale from $3.3398$ to $6.9690$~bps/Hz as the transmit SNR increases from $100$ to $110$~dB. Meanwhile, the other user rates increase gradually and stay above the minimum threshold. 

Consequently, the overall weighted sum rate (WSR) increases from $1.7150$ to $3.1953$~bps/Hz over the considered SNR range. For comparison, the dotted curves show the equal-weight baseline case $\boldsymbol{w}=[0.25, 0.25, 0.25, 0.25]$, where User~1 achieves the highest rate as the near user (NOMA 1 user), followed by Users~2, 3, and 4. This confirms that the user rate distribution is controlled flexibly by the weighting factors rather than being determined solely by the channel locations, enabling balanced NOMA resource control while satisfying the minimum QoS requirement. Furthermore, the evaluations showcase that an equal-weight priority assignment among NOMA users is sub-optimal in terms of maximizing the total WSR.

\begin{figure}[!t]
    \centering
    \includegraphics[width=8.5cm]{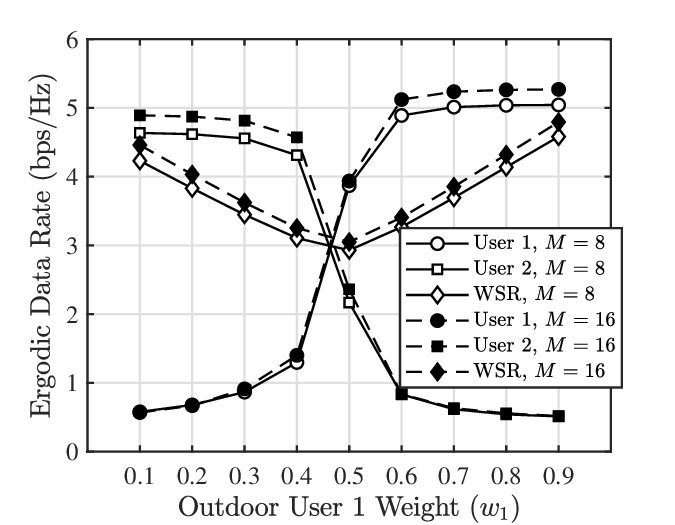}
    \caption{Achieved outdoor-user rates and total WSR versus for two users. The weight assigned to User~1 for $M=8$ and $M=16$ under $R^{\rm min}_{\rm in} =R^{\rm min}_{\rm out, k}=0.5$~bps/Hz for both indoor and outdoor users and $P/\sigma^2=100$~dB.}
    \label{fig:wsr_weight}
\end{figure}

Fig.~\ref{fig:wsr_weight} depicts individual user rates and the total WSR variations with priority weights for a $K=2$  (placed at $[10,12]$ m and uniformly in $[-\pi,\pi]$ from BS) scenario across $M=8$ and $M=16$ antennas. As $w_1$ increases from $0.1$ to $0.9$, the achieved rate of User~1 increases from $0.5777$ to $5.0417$~bps/Hz for $M=8$, whereas User~2 decreases from $4.6347$ to $0.5116$~bps/Hz. Similarly, for $M=16$, User~1's rate climbs from $0.5678$ to $5.2682$~bps/Hz while User~2's rate drops from $4.8905$ to $0.5166$~bps/Hz.  This confirms that the WSR formulation successfully redistributes resources according to user priorities. The WSR reaches its minimum at $w_1=0.5$ ($2.9278$ and $3.0481$~bps/Hz for $M=8$ and $M=16$), beyond which it increases to $4.5807$ and $4.7930$~bps/Hz at $w_1=0.9$. The consistently higher WSR obtained with $M=16$ validates the benefit of additional spatial degrees of freedom. Notably, the WSR is higher when priority is biased toward User~1 rather than User~2 due to User~1's proximity to the BS, which reduces large-scale propagation path loss and optimizes total power efficiency. Meanwhile, the QoS satisfaction probability remains close to 100\% across both configurations.

\begin{figure}[!t]
    \centering
    \includegraphics[width=8.5cm]{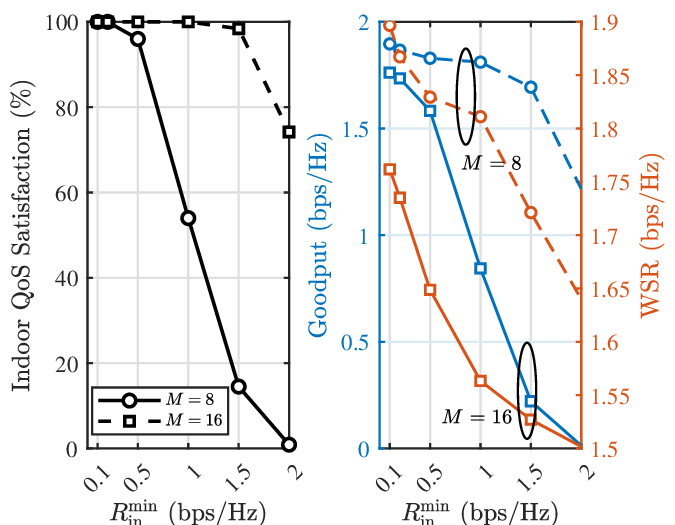}
    \caption{Indoor QoS reliability versus the indoor rate threshold for $K=4$ users placed at $[10, 12,15,18]$~m under a transmit power SNR of $100$~dB. Goodput=WSR$\times$ QoS probability.}
    \label{fig:QoS}
\end{figure}

The impact of the indoor rate requirement on outdoor NOMA performance and indoor QoS reliability is investigated in Fig.~\ref{fig:QoS}. For $M=8$, an outdoor threshold of $R_{\rm out,k}^{\min}= 0.5$~bps/Hz, and a transmit SNR of $100$~dB with $\mathbf{w}=[0.1, 0.3, 0.15, 0.25]$, increasing the indoor rate threshold from $R_{\rm in}^{\min}=0.1$ to $1$~bps/Hz reduces the outdoor WSR from $2.0724$ to $1.8300$~bps/Hz. User~2's rate drops from $3.6159$ to $3.1400$~bps/Hz, while Users~1, 3, and 4 track their QoS lines. The corresponding indoor QoS satisfaction probabilities are $100\%$, $100\%$, $96\%$, and $54\%$ for $R_{\rm in}^{\min}=\{0.1, 0.2, 0.5, 1\}$~bps/Hz, respectively, showing that stringent indoor bounds heavily reduce indoor QoS reliability.  With an alternative priority vector $\mathbf{w}=[0.1, 0.4, 0.3, 0.2]$, the outdoor WSR varies from $1.5884$ to $1.7800$~bps/Hz. For $M=16$ using $\mathbf{w}=[0.1, 0.4, 0.3, 0.2]$, the indoor QoS satisfaction probability remains a perfect $100\%$ for $R_{\rm in}^{\min}\leq1$~bps/Hz, dropping to $98.4\%$ and $74.2\%$ at $1.5$ and $2$~bps/Hz. Outdoor user rates stay highly stable across these thresholds, with User~2 achieving high rates between $3.1486$ and $3.7650$~bps/Hz. This comparison demonstrates that increasing the number of antennas provides essential spatial degrees of freedom, enabling the system to meet stringent indoor QoS requirements while maintaining highly stable outdoor NOMA performance.

\section{Conclusion}
This paper proposed an indoor--outdoor NOMA framework integrating AF relaying, SWIPT, pinching-antenna positioning, and ZF beamforming. The SWIPT-enabled relay uses harvested RF energy for indoor forwarding, while pinching-antenna positioning provides an additional degree of freedom for improving indoor coverage. A joint WSR maximization problem was formulated under indoor and outdoor QoS constraints. Due to its non-convexity, the problem was decomposed into indoor operating-point optimization and outdoor NOMA power allocation, solved using AO and SQP, respectively. Numerical results showed that the proposed framework effectively balances indoor QoS and outdoor WSR. Higher indoor rate requirements reduce outdoor performance, whereas additional BS antennas improve QoS satisfaction. The results demonstrate the potential of the proposed architecture for efficient indoor--outdoor wireless communications.
\bibliographystyle{IEEEtran}

\end{document}